\documentclass[]{uiuc_paper}
\pdfoutput=1

\usepackage[page,header]{appendix}
\usepackage{minitoc}
\usepackage{cleveref}
\crefname{figure}{Figure}{Figures}
\Crefname{figure}{Figure}{Figures}
\crefname{table}{Table}{Tables}
\Crefname{table}{Table}{Tables}
\crefname{section}{Section}{Sections}
\Crefname{section}{Section}{Sections}
\crefname{subsection}{Section}{Sections}
\Crefname{subsection}{Section}{Sections}
\crefname{equation}{Eq.}{Eqs.}
\Crefname{equation}{Equation}{Equations}
\usepackage{booktabs}
\usepackage{tabularx}
\usepackage{amsmath}
\usepackage{amssymb}
\usepackage{mathtools}
\usepackage{algorithm}
\usepackage{algorithmic}
\usepackage{colortbl}
\usepackage{multicol}
\usepackage{tikz}
\usepackage{pgfplots}
\pgfplotsset{compat=1.18}
\usetikzlibrary{positioning, arrows.meta, decorations.pathmorphing, calc, fit, backgrounds}
\usepackage{fix-cm}
\usepackage{lmodern}
\usepackage[T1]{fontenc}

\usepackage{latexsym}
\usepackage{url}
\usepackage{amsfonts}
\usepackage[utf8]{inputenc}
\usepackage{pifont}
\usepackage{multirow}
\usepackage{makecell}
\usepackage{paralist}
\usepackage{xspace}
\usepackage{adjustbox}

\newcommand{\tv}{$\tau$-Voice}
\newcommand{\ttb}{$\tau^2$-bench}
\newcommand{\tb}{$\tau$-bench}
\newcommand{\tc}[1]{\texttt{\small #1}}

\definecolor{pteal}{HTML}{0F6E8C}
\definecolor{lightblue}{RGB}{200,230,245}
\definecolor{lightgreen}{RGB}{180,245,180}

\newtcolorbox{insightbox}[1][]{
  enhanced,
  colback=teal!10,
  colframe=teal!60,
  fonttitle=\bfseries,
  coltitle=teal!80!black,
  colbacktitle=teal!30,
  title=#1,
  left=4mm, right=4mm, top=1mm, bottom=2mm,
  boxrule=1pt,
  arc=4mm,
  toptitle=2mm, bottomtitle=1mm
}

\newtcolorbox{findingbox}[1][]{
  enhanced,
  colback=orange!8,
  colframe=orange!30,
  fonttitle=\bfseries,
  coltitle=black,
  colbacktitle=orange!45,
  title=#1,
  left=4mm, right=4mm, top=1mm, bottom=2mm,
  boxrule=0.8pt,
  arc=3mm,
  toptitle=2mm, bottomtitle=1mm
}

\newcommand{\method}[0]{{\usefont{T1}{lmr}{m}{sc}SpeechGym}\xspace}

\title{\method: An Audio-Native Gym for Training Voice Agents via Reinforcement Learning}

\author[1,2]{Jiajun Fan}
\author[2]{Jingyuan Li}
\author[2]{Prashanth Gurunath Shivakumar}
\author[2]{Jia-Hong Huang}
\author[2]{Qi Luo}
\author[2]{M. Maruf}
\author[2]{Ivan Bulyko}
\author[1]{Ge Liu}
\author[2]{Roger Ren}

\affiliation[1]{University of Illinois Urbana-Champaign}
\affiliation[2]{Amazon AGI Foundations}

\abstract{Voice agents must call tools and hold multi-turn dialogue entirely through speech, yet the
dominant paradigm trains them in text. Existing frameworks either cascade TTS and ASR around a
proprietary voice API, where gradients cannot flow and per-call cost makes on-policy
reinforcement learning prohibitive, or stay in text: they measure voice agents but cannot
improve them. We present SpeechGym, an audio-native agentic environment in which two
omni-modal models converse in native audio, with no external ASR or TTS and no API boundary,
over the unmodified tasks, tools and success check of an established text agentic benchmark,
so that the interaction modality is the only variable and the loop stays local and trainable
end to end. Audio agentic capability does not follow from audio understanding. The failures
speech introduces are perceptual rather than reasoning deficits: the agent picks the right tool
and the right argument slot but fills it with a value misheard from the waveform, and that
single error cascades into a failed call, a retry of the same call, and a wasted step budget.
A second failure is behavioural: under an insistent caller the agent performs an unauthorised
write and ends the episode believing it helped. Both are trainable, because the environment
labels them for free: a call with a misheard argument fails against the database while a
correct one succeeds. The obstacle is sparsity, not signal. Outcome-only GRPO is
gradient-starved here, since almost every rollout group fails identically, while a per-turn
process reward crediting each successful tool call restores variance to nearly every group.
Trained this way, the agent transfers with no further tuning to an independently implemented
voice benchmark, more than doubling task success and carrying an open-weights model from last
place to second on that leaderboard, while using fewer turns and tokens than before training.
}

\date{\today}

\begin{document}

\maketitle

\section{Introduction}\label{sec:intro}

A voice agent that changes a booking or disputes a charge must do everything a text agent
does---call tools against a live database, respect a domain policy, drive a multi-turn
dialogue to a verifiable end state---with speech as its only channel. The dominant recipe
trains the policy in text and attaches speech at the edges, assuming competence acquired in
text survives the round trip through audio. Omni-modal models
\citep{xu2025qwen25omni,qwen2025qwen3omni,defossez2024moshi,fu2024vita,xie2024miniomni}
dissolve the cascade architecturally but supply no way to \emph{train} in that regime: RL on
them has addressed only single-turn tasks with no tools, no dialogue partner and no
environment state \citep{rouditchenko2025omnir1,fancesar,zhao2025keomnir}, while agentic RL
is almost entirely textual \citep{qian2025toolrl,zheng2026procedureaware,modecrua2026multiturn}. How to train an agent
natively in audio, and what breaks once the text safety net goes, remains open.

\tv{} \citep{ray2026tauvoice} makes the gap concrete by wrapping \ttb{}
\citep{barres2025tau2}, itself an extension of \tb{} \citep{yao2024taubench}, in a voice loop:
a user LLM writes the caller's turn, a TTS service speaks it, and the agent is a proprietary
real-time API. It reports a steep text-to-speech drop, but cannot close the gap it exposes:
gradients do not flow through a closed API, and its latency and price rule out the rollout
volume on-policy RL needs (\cref{sec:related}). Voice-agent benchmarks
\citep{chen2024voicebench,wang2024audiobench,jain2025voiceagentbench,lin2026fullduplex} share
the property: voice agents can be measured, not improved.

\begin{figure*}[t]
\centering
\definecolor{pkF}{HTML}{FBE9EC}\definecolor{pkB}{HTML}{C48F99}
\definecolor{lvF}{HTML}{E9E9F6}\definecolor{lvB}{HTML}{8F8FBE}
\definecolor{mtF}{HTML}{E4F2EC}\definecolor{mtB}{HTML}{7FB79E}
\definecolor{crF}{HTML}{FBF3E4}\definecolor{crB}{HTML}{CFB183}
\definecolor{arB}{HTML}{6B76C9}\definecolor{arT}{HTML}{4E9C86}
\definecolor{arO}{HTML}{DFA046}\definecolor{txN}{HTML}{2F3542}
\definecolor{txG}{HTML}{7A8290}

\resizebox{0.9\linewidth}{!}{%
\begin{tikzpicture}[font=\normalsize, text=txN,
  bx/.style={rounded corners=3pt, line width=0.7pt, minimum height=13mm, align=center},
  usr/.style={bx, draw=pkB, fill=pkF, minimum width=36mm},
  agt/.style={bx, draw=lvB, fill=lvF, minimum width=36mm},
  tls/.style={bx, draw=mtB, fill=mtF, minimum width=38mm},
  env/.style={bx, draw=mtB, fill=mtF, minimum width=40mm},
  rl/.style={bx, draw=crB, fill=crF, minimum width=42mm, minimum height=11mm},
  ar/.style={-{Stealth[length=2.6mm]}, line width=0.9pt},
  lbl/.style={font=\small\itshape, text=txG}]

\node[font=\bfseries, anchor=west] at (-0.6,2.0)
  {(a) A speech agent runs the same agentic loop --- the only variable is the channel};
\node[usr] (u) at (0,0)    {\textbf{User}\\[1pt]\small real person \emph{or} model};
\node[agt] (a) at (6.2,0)  {\textbf{Agent}\\[1pt]\small LLM};
\node[tls] (t) at (12.6,0) {\textbf{Tools + DB}\\[1pt]\small book, cancel, look up records};

\draw[ar, arB] (u.north) to[bend left=20]
  node[above, font=\small\bfseries, text=arB] {communication: text \emph{or} speech} (a.north);
\draw[ar, arO] (a.south) to[bend left=20] node[below, lbl] {reply (typed / spoken)} (u.south);
\draw[ar, arT, dashed] ([yshift=3mm]a.east)  -- node[above, lbl] {tool call} ([yshift=3mm]t.west);
\draw[ar, arT, dashed] ([yshift=-3mm]t.west) -- node[below, lbl] {result}    ([yshift=-3mm]a.east);
\node[below=4mm of t, lbl] {same tools either way};

\begin{scope}[yshift=-4.6cm]
  \node[font=\bfseries, anchor=west] at (-0.6,2.0)
    {(b) SpeechGym closes that loop locally --- which is what makes the agent trainable};
  \node[agt] (sa) at (0.8,0)   {\textbf{Speech Agent}\\[1pt]\small (trainable)};
  \node[env] (ev) at (8.6,0)   {\textbf{Environment}\\[1pt]\small (User + Tools + DB)};
  \node[rl]  (up) at (4.7,-2.8){\textbf{RL Update (GRPO)}};

  \draw[ar, arB] ([yshift=3mm]sa.east) --
    node[above, font=\small\bfseries, text=arB] {speech + tool calls} ([yshift=3mm]ev.west);
  \draw[ar, arO] ([yshift=-3mm]ev.west) -- node[below, lbl] {speech + tool results} ([yshift=-3mm]sa.east);
  \draw[ar, arT] (ev.south) |- node[above, pos=0.72, lbl, text=arT] {reward} (up.east);
  \draw[ar, arO] (up.west) -| node[below, pos=0.28, lbl, text=arO] {update $\theta$} (sa.south);
\end{scope}
\end{tikzpicture}}

\caption{\textbf{\method overview.} \textbf{(a)}~A speech agent runs the same agentic loop as a
text agent: it talks to a user, calls tools against a live database, and is scored on the final
state. The one thing that changes is the channel carrying the conversation --- which is why
holding tasks, tools and the success check fixed makes modality a controlled variable.
\textbf{(b)}~SpeechGym closes that loop locally. The user simulator, the tools and the reward
all live inside the environment, so speech rollouts can be scored and turned into a policy
update rather than merely measured. No external ASR or TTS and no proprietary API sits in the
loop, which is what makes gradients --- and therefore training --- possible at all.}
\label{fig:architecture}
\end{figure*}
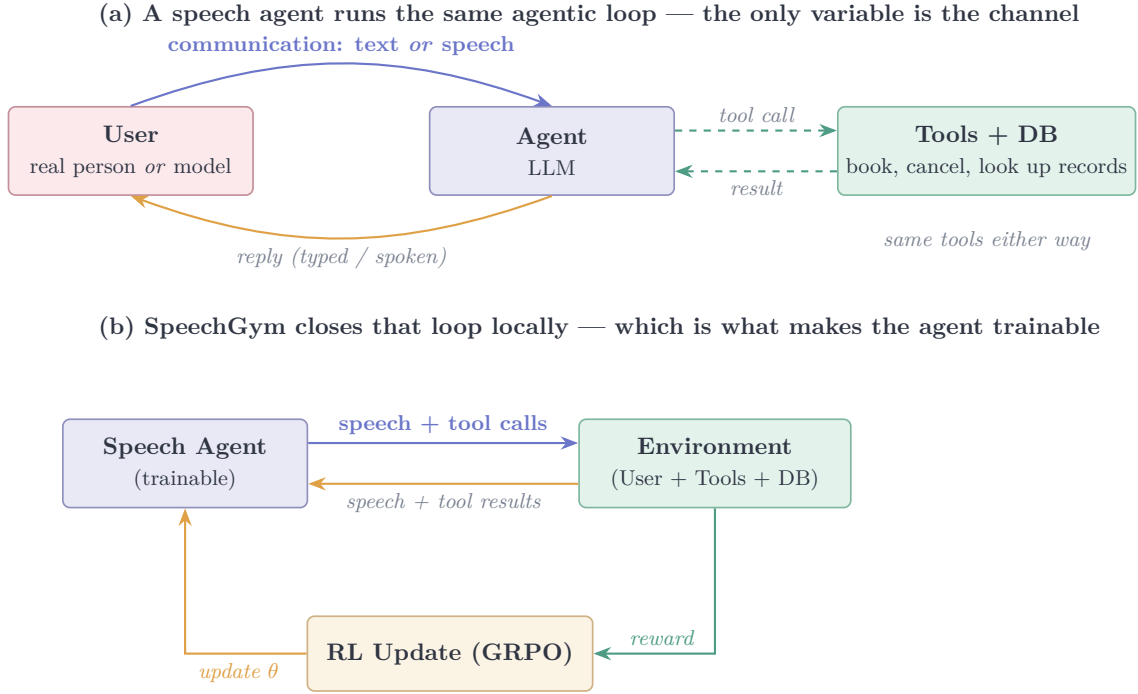

We introduce \textbf{SpeechGym} (\cref{fig:architecture}), an audio-native agentic environment
whose loop is entirely local and therefore trainable. Two omni-modal models converse in native
audio: a frozen user model $\pi_U$ speaks the caller's side, and the trainable
Thinker--Talker agent $\pi_\theta$ chooses at every step between a structured tool call,
executed deterministically against the database, and speaking back to the user. With no
external ASR, TTS or API boundary in the loop, rollouts cost only local compute and the policy
stays ours. Everything but the interaction modality is inherited unmodified from
the underlying text benchmark---tasks, tools, databases, policies, success check---so any
difference in success is attributable to modality alone.

In general, our contributions are as follows:
\begin{itemize}
\item \textbf{An audio-native agentic gym.} SpeechGym is, to our knowledge, the first
environment to combine audio-native interaction, multi-turn tool use and end-to-end RL
trainability; prior frameworks supply at most two, and every prior audio framework is
evaluation-only because an API cascade on the user side can be evaluated but never trained
(\cref{sec:related}). It exposes a reset/step/reward interface that accepts any text agentic domain of this
shape and any open omni-modal model, and adds a Banking domain stressing high-stakes numeric slots
(\cref{sec:method}).

\item \textbf{A diagnosis of the text-to-audio gap.} The dominant failures are perceptual, not
reasoning deficits: the base 30B omni model mis-hears a slot value in 32\% of speech rollouts
against 2\% in text, a sixteen-fold increase that cascades into a 42\% tool-error rate and
into dead loops ending the episode at zero reward. Alongside it sits confidently-wrong
over-action, observed throughout the speech rollouts but not quantified against text
(\cref{tab:failures}, \cref{sec:failures}).

\item \textbf{A training recipe for the low-base-rate regime.} Because GRPO
\citep{shao2024deepseekmath,guo2025deepseekr1} normalises returns within a group, near-floor
audio success rates leave only 16\% of groups carrying any gradient; a per-turn process reward,
dense shaping in the classical sense \citep{ng1999shaping}, raises this to 99.6\%
(\cref{tab:groups}, \cref{sec:training}). A vLLM-based \citep{kwon2023vllm} omni-modal rollout
server makes a training epoch 5.4$\times$ faster at \$0 API cost.

\item \textbf{Cross-pipeline transfer.} Run with no further tuning on \tv{}'s independently
implemented pipeline---a different user simulator, TTS and ASR stack, and evaluation
harness---the trained agent more than doubles pass@1, from 24\% to 53\%, with gains in all
three domains and non-overlapping confidence intervals (\cref{fig:tauvoice},
\cref{sec:transfer}), moving an open 30B model from last place to second on that
leaderboard, ahead of the cascaded baseline and of proprietary real-time systems as reported
by their providers and not re-run by us (\cref{fig:leaderboard}). The agent also uses
fewer turns and tokens (\cref{tab:efficiency}), issues fewer unauthorised writes, falls into
fewer dead loops, and recovers from mis-hearings more often (\cref{tab:behaviour}).
\end{itemize}

The text-to-audio gap is therefore not a fixed cost of the modality but a deficit in
identifiable competences---hearing a slot value correctly, confirming before acting,
abandoning a failing plan---that closed-loop training substantially reduces.

\section{Related Work}\label{sec:related}

\subsection{Tool-using language agents}
Tool augmentation makes a language model an agent that acts on external state
\citep{schick2023toolformer,qin2024toolllm}; \tb{} \citep{yao2024taubench} made this rigorous:
customer service as a POMDP with a simulated user, documented database tools and an automatic
final-state check. \ttb{} \citep{barres2025tau2} generalises it to a dual-control Dec-POMDP where
the user also holds tools. Web navigation \citep{zhou2024webarena}, software
engineering \citep{jimenez2024swebench} and executable multi-hop retrieval
\citep{pyrag} stress other axes but remain textual. Standardised RL
environments turn a capability into a trainable, reproducible target: the Arcade Learning
Environment \citep{bellemare2013ale} sustained a decade of algorithmic work that eventually
pushed past human world records \citep{gdi,lbc}. SpeechGym plays that role for voice agents,
keeping \tb{}'s evaluation methodology while making the channel audio-native and the loop
trainable.

\subsection{Voice agent evaluation}
\tv{} \citep{ray2026tauvoice} is the closest prior effort: it extends \ttb{} to voice, driving
the caller's side through a cascade---a text user simulator feeding a TTS system---while the
agent is a proprietary full-duplex realtime voice API. It measures rather than trains: no
gradient flows through a proprietary endpoint, and the cascade's latency and price preclude
the rollout volume on-policy RL needs---one epoch costs over \$200 in API calls, a
seven-epoch run over \$1{,}400, and thousand-epoch budgets approach \$200k for a
single run. SpeechGym runs the same loop locally at \$0 API cost. Other audio benchmarks share
the scope: VoiceAgentBench \citep{jain2025voiceagentbench} and Full-Duplex-Bench
\citep{lin2026fullduplex} assess spoken tool use without training; VoiceBench
\citep{chen2024voicebench} and AudioBench \citep{wang2024audiobench} test single-turn spoken
understanding without tools.

\subsection{Omni-modal models}
Qwen2.5-Omni \citep{xu2025qwen25omni} introduces the Thinker--Talker design, whose hidden
states drive an autoregressive speech-token decoder; Qwen3-Omni \citep{qwen2025qwen3omni}
scales it with a Mixture-of-Experts backbone, and VITA \citep{fu2024vita}, Mini-Omni
\citep{xie2024miniomni} and Moshi \citep{defossez2024moshi} pursue open speech-to-speech
interaction. They understand audio and generate speech, but none is trained for audio
\emph{agentic} tasks: when to invoke a tool rather than speak, when to confirm a misheard
value, when a write is authorised. SpeechGym supplies that signal.

\subsection{RL for reasoning and tool use}
RL with verifiable rewards is now standard for eliciting reasoning, in general domains
\citep{guo2025deepseekr1} and in mathematics \citep{yang2026batched}.
We optimise with GRPO \citep{shao2024deepseekmath}, a group-relative alternative to PPO's
learned value function \citep{schulman2017ppo}, with per-turn shaping in the classical
sparse-reward sense \citep{ng1999shaping}. In audio, RL has so far targeted reasoning
over audio inputs \citep{zhao2025keomnir,rouditchenko2025omnir1}, including with process-level
rewards over the reasoning trace \citep{fancesar}; in all of these an episode is a single
question and the model never acts on external state. In text, RL for tool use has largely
optimised a single call per episode, whether through reward design \citep{qian2025toolrl} or
procedure-aware supervision of the call \citep{zheng2026procedureaware}; closest on the task
side, multi-turn GRPO has been applied to \tb{}'s text mode \citep{modecrua2026multiturn}. Outside language, post-training and adaptive
computation for large multimodal policies are driven by the same constraint we face --- the
cost of acting, not of updating --- whether by scheduling and pruning a
vision-language-action model \citep{li2026spvla}, learning when to deliberate at all
\citep{elegantvla}, or adapting inference structure to the input \citep{prance}.
To our knowledge, no prior work applies RL to audio \emph{agentic} tasks, where perception,
dialogue and tool use are optimised jointly.

\subsection{Positioning}
Three axes matter---audio-native interaction, multi-turn tool use, trainability by online
RL---and prior work supplies at most two: \tb{}, \ttb{}
\citep{yao2024taubench,barres2025tau2} and single-call tool-use RL
\citep{qian2025toolrl,zheng2026procedureaware} are trainable but text-only, and every prior audio framework
\citep{ray2026tauvoice,jain2025voiceagentbench,lin2026fullduplex,chen2024voicebench,wang2024audiobench}
is evaluation-only. The decisive difference is the user side: an API cascade can be measured
but never differentiated through, and its per-rollout cost rules out on-policy training. The
claim needs care: the text benchmarks' public code releases do expose a Gymnasium-compatible
interface, so their \emph{text} domains can in principle be trained in, even though they are
presented and used as evaluation suites. No such route exists for the voice setting: there the
agent is a proprietary realtime endpoint and the caller is synthesised by a commercial TTS
service, so the audio loop admits no gradient at either end. SpeechGym is, to our knowledge,
the only system simultaneously audio-native, multi-turn, tool-using and trainable end to end
by RL, because both sides of the conversation are local open models inside the agent's loop.
The roles are complementary: we train in SpeechGym and evaluate, untuned, on \tv{}'s
pipeline and scoring code (\cref{sec:transfer}).

\section{SpeechGym}\label{sec:method}

SpeechGym takes a text agentic benchmark and makes it audio-native and trainable, leaving the
task definition alone: tasks, tools, databases and the success check are inherited unmodified,
and only the channel between user and agent changes, from text to native speech. We
instantiate it on \ttb{} \citep{barres2025tau2} throughout, but nothing in the design is
specific to that suite.

\subsection{Problem formulation}\label{sec:pomdp}

An audio agentic episode is a partially observable Markov decision process
$(\mathcal{S}, \mathcal{A}, \mathcal{O}, P, R)$ with machine-checkable elements. The state
$s \in \mathcal{S}$ is the hidden domain database --- customer records, reservations, order
lines, account balances --- never observed directly and never described in the context: it
is read only through a documented read tool and changed only by a write tool invoked with
correct arguments.

An observation $o \in \mathcal{O}$ is a user audio waveform from the frozen user model or the
textual result of an executed tool. An action $a \in \mathcal{A}$ is a tool call
$a_{\text{tool}} = (\texttt{name}, \texttt{args})$, emitted as structured text and dispatched
to the executor, or a variable-length speech response
$a_{\text{speech}} \in \mathbb{R}^{L}$ synthesised as a waveform; which one it emits is part
of the policy's decision at every step (\cref{sec:arch}).

Tool calls transition $s$ deterministically: read tools return records without changing it,
write tools may change it, and invalid or unauthorised calls return an error message leaving
it untouched. A speech action instead conditions the frozen user model, which replies in
audio. The episode ends on the user's resolution signal or at the step budget
$T_{\max} = 50$.

This is the text-mode POMDP of \tb{} \citep{yao2024taubench} with raw audio in place of the
user's transcript, which introduces three difficulties: slot values must be extracted from a
waveform rather than copied from a string, so one mis-perceived character propagates into a
tool argument; prosody carries urgency, hesitation and insistence, which bias the decision to
act; and the policy must alternate between tool-call syntax and free spoken language within
one context.

\subsection{Environment architecture}\label{sec:arch}

\begin{figure}[t]
\centering
\definecolor{ckF}{HTML}{FBE9EC}\definecolor{ckB}{HTML}{C48F99}
\definecolor{cvF}{HTML}{E9E9F6}\definecolor{cvB}{HTML}{8F8FBE}
\definecolor{cmF}{HTML}{E4F2EC}\definecolor{cmB}{HTML}{7FB79E}
\definecolor{caB}{HTML}{6B76C9}\definecolor{caT}{HTML}{4E9C86}
\definecolor{caO}{HTML}{DFA046}\definecolor{ctN}{HTML}{2F3542}
\definecolor{ctG}{HTML}{7A8290}

\resizebox{\linewidth}{!}{%
\begin{tikzpicture}[font=\normalsize, text=ctN,
  bx/.style={rounded corners=3pt, line width=0.7pt, minimum height=17mm, align=center},
  usr/.style={bx, draw=ckB, fill=ckF, minimum width=40mm},
  agt/.style={bx, draw=cvB, fill=cvF, minimum width=40mm},
  env/.style={bx, draw=cmB, fill=cmF, minimum width=40mm},
  ar/.style={-{Stealth[length=2.6mm]}, line width=0.9pt},
  lbl/.style={font=\small\itshape, text=ctG}]

\node[usr] (u) at (0,0)      {\textbf{User Model} $\pi_U$\\[1pt]\small Qwen3-Omni-30B\\[1pt]\small \emph{frozen}};
\node[agt] (a) at (7.4,0)    {\textbf{Agent} $\pi_\theta$\\[1pt]\small Qwen3-Omni-30B\\[1pt]\small \emph{trained}};
\node[env] (t) at (14.4,0)   {\textbf{Tool Executor} $\mathcal{E}$\\[1pt]\small tools $+$ database};
\node[env] (r) at (14.4,-4.0){\textbf{Reward} $R$\\[1pt]\small task-completion check\\[1pt]\small (automatic)};

\draw[ar, caB] (u.north) to[bend left=20] node[above, lbl] {speech} (a.north);
\draw[ar, caO] (a.south) to[bend left=20] node[below, lbl] {spoken reply} (u.south);
\draw[ar, caT, dashed] ([yshift=3.5mm]a.east)  -- node[above, lbl] {tool call} ([yshift=3.5mm]t.west);
\draw[ar, caT, dashed] ([yshift=-3.5mm]t.west) -- node[below, lbl] {result}    ([yshift=-3.5mm]a.east);
\draw[ar, caT] (t.south) -- (r.north);
\end{tikzpicture}}

\caption{\textbf{The four components, as we instantiate them.} The \emph{same} omni-modal
model plays both sides: one frozen copy speaks the caller's side in native audio, while the
trainable copy decides at every step between emitting a structured tool call and speaking back. The executor runs the tools against the database, and the reward is the
benchmark's own task-completion check, computed automatically. Only the user--agent channel is
audio: keeping the tool interface textual separates \emph{perceptual} error --- mishearing a
slot value --- from \emph{behavioural} error --- choosing the wrong tool or acting without
authorisation.}
\label{fig:components}
\end{figure}
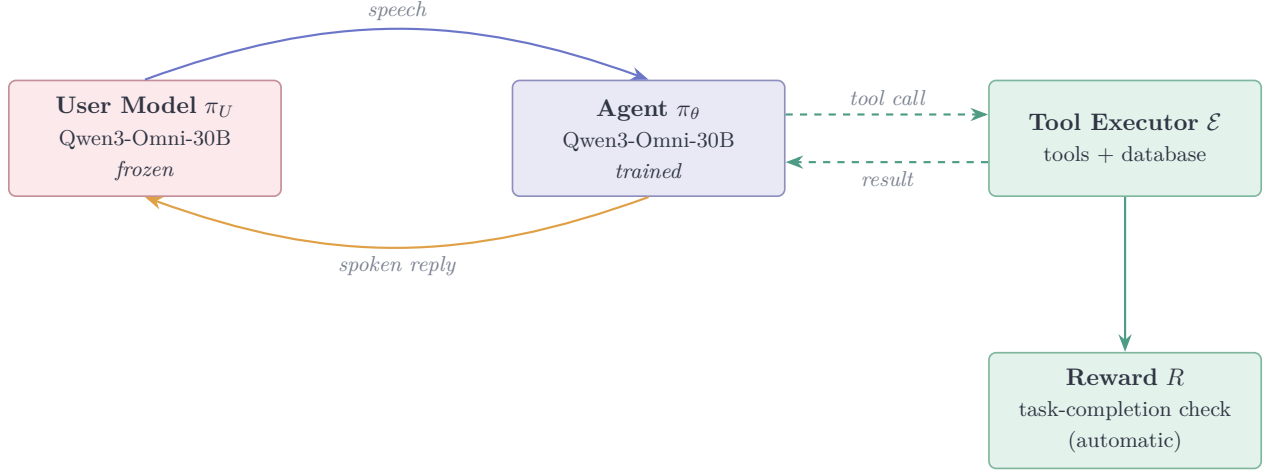

SpeechGym has four components (\cref{fig:components}).

\textbf{User model $\pi_U$.} A frozen Qwen3-Omni-30B-A3B \citep{qwen2025qwen3omni}, given the
task scenario (persona, goal, what the user knows), generates its turns directly as native
audio, with no intermediate text-to-speech stage. Never updated, it fixes the speech
distribution against which policy improvement is measured.

\textbf{Agent model $\pi_\theta$.} The trainable policy: the same Thinker--Talker omni-modal
model, Qwen3-Omni-30B-A3B \citep{xu2025qwen25omni,qwen2025qwen3omni}. The Thinker consumes
the audio and tool observations, reasons in text, and \emph{autonomously} decides the action
type: output containing a tool-call pattern is parsed and dispatched, otherwise the Talker
synthesises it into speech. No external controller or scripted schedule governs that choice.

\textbf{Tool executor $\mathcal{E}$.} The benchmark's own executor and databases, unmodified. Failed
calls return an error message and leave the state unchanged.

\textbf{Evaluator.} The benchmark's own outcome check, run unchanged (\cref{sec:reward}).

\textbf{Why tool calls stay textual.} Only the user--agent channel is audio; tool calls and
results stay structured text. This separates \emph{perceptual} error --- mishearing the
caller and writing a wrong slot value --- from \emph{behavioural} error --- wrong tool,
skipped step, action without authorisation: over a noiseless tool interface a wrong argument
was misheard or mis-reasoned, not corrupted in transit. Tool calls also execute instantly and
do \emph{not} consume a conversational turn, so the agent may chain several before speaking,
whereas a speech action always advances the dialogue with $\pi_U$.

\textbf{Domains.} The three \ttb{} domains --- Airline, Retail and Telecom, the last
dual-control, where the user also holds tools the agent must talk them through --- plus
Banking, for high-stakes numeric slots. Banking is built like a \ttb{} domain: accounts,
balances and transactions in a relational database; read tools for lookup, write tools for
transfers, disputes and limit changes; per-task gold action sequences verified in text mode;
and a final database-state check. \tv{} \citep{ray2026tauvoice} covers only the three
original domains, hence Banking's absence from \cref{sec:transfer}.

\textbf{Trainability.} With no intermediate ASR or TTS and both models open and local, the
loop is differentiable at the agent and free to sample from; behind a proprietary voice API
gradients cannot cross the boundary, and latency and price cap rollouts (\cref{sec:system}).

\subsection{Reward}\label{sec:reward}

We do not design a reward: the episode outcome is scored by \ttb{}'s own evaluator, run
unchanged, so a SpeechGym reward and a \ttb{} score mean the same thing. The evaluator
defines five binary components, of which each task selects a subset --- its reward basis
$\mathcal{B}(\tau)$:

\begin{description}\setlength{\itemsep}{1pt}
  \item[$r_{\text{DB}}$] the gold action sequence is replayed on a fresh environment and the
  resulting database compared with the agent's; a match scores $1$.
  \item[$r_{\text{COMM}}$] the replies contain every information string the task requires,
  such as a confirmation number.
  \item[$r_{\text{ACTION}}$] the required write tools were called with correct arguments.
  \item[$r_{\text{ENV}}$] the user's device reaches the expected final state, for example
  data enabled --- the dual-control criterion.
  \item[$r_{\text{NL}}$] the task's natural-language assertions hold, for example that the
  agent confirmed the refund.
\end{description}

The episode reward is the product over that basis,
\begin{equation}\label{eq:reward}
R(\tau) \;=\; \prod_{c \,\in\, \mathcal{B}(\tau)} r_c, \qquad r_c \in \{0,1\},
\end{equation}
so every required condition must hold; there is no partial credit, and an episode that
exhausts \tc{max\_steps} scores $R = 0$.

\cref{eq:reward} is identical in training and evaluation, and modality-independent: it reads
database rows, tool arguments and required strings, never how the interaction was conducted.
The only training-time addition is the shaping of \cref{sec:process}, which leaves the
terminal outcome untouched. Reward parity was audited: the SpeechGym reward path reproduces
\ttb{}'s scores exactly on golden trajectories, and the vLLM and non-vLLM code paths agree.

\subsection{Training with GRPO}\label{sec:grpo}

For each task $\tau_i$ we sample $K = 4$ complete speech episodes under the current policy:
full multi-turn conversations with the frozen user model interleaved with tool executions,
terminating by user signal or at $T_{\max}$. GRPO \citep{shao2024deepseekmath} replaces
PPO's value function \citep{schulman2017ppo} with a group baseline: with $r_{i,j}$ the return
of the $j$-th rollout of $\tau_i$ and $\mu_i, \sigma_i$ the group's mean and standard
deviation,
\begin{equation}\label{eq:adv}
\hat{A}_{i,j} \;=\; \frac{r_{i,j} - \mu_i}{\sigma_i + \epsilon},
\end{equation}
so a group with identical returns has $\sigma_i = 0$ and no gradient (\cref{sec:process}). In
the process variant the same normalisation is applied per turn: $G_t$ is standardised by the
group statistics and broadcast to the turn's tokens. The update optimises
\begin{equation}\label{eq:loss}
\mathcal{L}(\theta) \;=\; -\sum_{i,j} \min\!\big(\rho_{i,j}\hat{A}_{i,j},\;
\mathrm{clip}(\rho_{i,j}, 1-\varepsilon, 1+\varepsilon)\,\hat{A}_{i,j}\big)
\;+\; \beta\, D_{\mathrm{KL}}\big[\pi_\theta \,\|\, \pi_{\text{ref}}\big],
\end{equation}
with $\rho_{i,j}$ the importance ratio to the policy that generated the rollout. How tightly to hold a fine-tuned policy to its
reference is an active design axis in RL post-training of generative models, from adaptive
divergence regularisation \citep{fan2025adaptive} to reward-weighted objectives with transport
regularisation \citep{fan2025online}. We adopt the KL-free variant, $\beta = 0$, relying on the
clip term and the adapter's low-rank constraint to stay near $\pi_{\text{ref}}$. Only a LoRA adapter
\citep{hu2022lora} is trained: rank $8$, $\alpha = 16$, on the linear projections of the
Thinker of a Mixture-of-Experts backbone. The speech-synthesis path is untouched: credit
assignment is scoped to \emph{what the agent does}, not \emph{how it sounds}.

\subsection{Densifying the reward: per-turn process shaping}\label{sec:process}

\cref{eq:reward} is computed once, at termination, and audio agentic tasks sit in a
low-base-rate regime (\cref{sec:failures}). When almost every episode fails, almost every
group of $K$ rollouts fails \emph{identically}: all returns are zero, $\sigma_i = 0$ in
\cref{eq:adv}, the group yields no gradient, and most rollout compute produces no learning
signal.

Following shaping practice \citep{ng1999shaping,modecrua2026multiturn}, we credit progress
per turn. Each successful tool execution receives $+0.1$ and each failed call $-0.1$; the
outcome $R(\tau)$ is appended to the last step, so the terminal criterion is preserved.
Per-turn returns are discounted sums,
\begin{equation}\label{eq:return}
G_t \;=\; \sum_{k=t}^{T} \gamma^{\,k-t} r_k, \qquad \gamma = 0.99,
\end{equation}
and replace the flat episode reward in \cref{eq:adv}. Failing rollouts can now differ in how
far they got, so variance appears in groups where none succeeded: four failing rollouts give
$\{0, 0, 0, 0\}$ under outcome-only reward and no gradient, whereas under process shaping the
same group might give $\{0.30, -0.10, 0.20, 0.10\}$, with
$\sigma_i > 0$ and a gradient toward the rollout that executed more tools successfully.
\cref{tab:groups} reports the effect on the fraction of groups that carry gradient.

\subsection{System: making online speech RL affordable}\label{sec:system}

Rollout collection, not the gradient step, is the bottleneck: an episode is dozens of turns,
each generating audio from two 30B models. We serve both with vLLM-Omni \citep{kwon2023vllm}:
the rollout worker calls an OpenAI-compatible endpoint with \tc{modalities:[text,audio]},
user audio passed as base64 WAV and the agent's speech returned alongside its text, so a turn
is one request rather than a chain of conversions. The trained LoRA is served per request by
adapter name, so an updated policy reaches the workers without reloading a merged checkpoint.

\textbf{Deployment.} One 8$\times$H200 pod hosts the loop: GPUs 0--1, 2--3 and 4--5 run three
vLLM-Omni servers, each hosting both models and using two GPUs for the reasoning and
speech-synthesis paths; GPUs 6--7 run LoRA training in bf16. A group's $K = 4$ rollouts run
on four threads round-robin across the three servers.

\textbf{Effect.} Multi-turn speech rollouts, not the policy update, dominate the epoch, so
serving them efficiently is what makes online speech RL practical: end to end, a training
epoch becomes $5.4\times$ faster.
An API cascade running the same rollouts has a per-epoch price that puts a full training run
out of reach (\cref{sec:related}); SpeechGym's training loop has \$0 API cost.

\textbf{Near-on-policy sampling.} The first group of a run is collected on base weights, no
adapter existing yet, and a group in flight when an update lands finishes under the previous
adapter, so rollouts are \emph{near}-on-policy. The clipped importance ratio in
\cref{eq:loss} is designed to tolerate this lag, but the deviation is real.

\section{Experiments}\label{sec:exp}

We ask what breaks when an agentic task moves from text to speech (\cref{sec:failures}),
whether the reward signal inside SpeechGym is dense enough to train on
(\cref{sec:training}), whether the result
survives outside the training environment (\cref{sec:transfer}), and why
(\cref{sec:mechanism}).

\subsection{Setup}\label{sec:setup}

\paragraph{Models.} Agent and frozen user simulator are both Qwen3-Omni-30B-A3B
\citep{qwen2025qwen3omni}, an omni-modal mixture-of-experts model with native audio input and
output. Only the agent is updated, through a LoRA adapter \citep{hu2022lora} on the Thinker
(\cref{sec:arch}, \cref{sec:grpo}); the user is never trained, making it a fixed --- if
idealised --- speech distribution. One model family on both sides keeps the loop local and
trainable (\cref{sec:system}); these results are reported in that setting.

\paragraph{Compute.} All runs use one 8$\times$H200 pod with vLLM-Omni rollouts, which makes a
training epoch $5.4\times$ faster end to end at \$0 API cost since both models are local.

\paragraph{Domains.} Airline, Retail and the dual-control Telecom domain, inherited unmodified
from \ttb{} \citep{barres2025tau2}, plus Banking, added for long, high-stakes numeric slots.
Tasks, tools, databases and the success check are \ttb{}'s, so the only variable between text
and speech runs is the channel.

\paragraph{Metric.} pass@1 under \ttb{}'s unchanged task-completion check: an episode succeeds
only if the final database state matches the state produced by replaying the gold actions
\emph{and} all required information has reached the caller. \tv{} \citep{ray2026tauvoice} uses
the same criterion, so our in- and out-of-environment numbers are comparable. The check reads
structured outcome fields only, rewarding neither fluent nor awkward-sounding speech.

\paragraph{Two evaluation axes.} \emph{In-gym} results, against the training-time user
simulator and clean self-play audio, are diagnostic only; the headline result is measured
entirely \emph{outside} the training environment, on \tv{} (\cref{sec:transfer}).

\subsection{What breaks in speech}\label{sec:failures}

\begin{table}[t]
\centering
\caption{\textbf{Failure modes amplified by speech.} Fraction of SpeechGym rollouts
exhibiting each failure under the two channels; same tasks, same tools, same reward.
Every mode is amplified by speech --- mis-hearing is near-absent in text, and the downstream
tool errors and dead loops are markedly rarer --- which is why a speech-native environment is
needed to expose them, and to supply the learning signal that fixes them.}
\label{tab:failures}
\small
\begin{tabular}{@{}lcccl@{}}
\toprule
\textbf{Failure type} & \textbf{Speech} & \textbf{Text} & \textbf{Ratio} & \textbf{Interpretation} \\
\midrule
Mis-hearing (slot value) & 32\% & 2\%  & 16$\times$  & hears the wrong name / ID / digit \\
Tool error rate          & 42\% & 26\% & 1.6$\times$ & mis-heard arguments $\rightarrow$ downstream errors \\
Dead loop                & 29\% & 18\% & 1.6$\times$ & retries the same failing call, no recovery \\
\bottomrule
\end{tabular}
\end{table}

We annotated rollouts of the same tasks, tools and reward in the two channels; only the
channel differs. Mis-heard slot values appear in 32\% of speech rollouts against 2\% in text,
a $16\times$ amplification; tool errors rise from 26\% to 42\%, dead loops from 18\% to 29\%
(\cref{tab:failures}). Three patterns dominate.

\paragraph{Slot-value extraction from audio.} The agent picks the right tool and the right
argument slot, then fills it with a mis-heard value: a digit of a zip code, a character of an
order ID, a spelling of a name. Plan and execution are correct; only the perceived value is
wrong --- a perceptual, not a reasoning, failure, and the largest gap between the channels.
The database check is exact, so one confused character is worth the same as no attempt.

\paragraph{Confidently-wrong over-action.} The agent performs a state-changing write it was
not authorised to perform and ends the episode believing it has helped. We report this
qualitatively, as it has no matched text baseline in \cref{tab:failures}: an insistent,
emotional caller tone is far more vivid in audio, and the agent concedes to pressure a
transcript would have flattened. Where the correct resolution is to decline or escalate, the
write corrupts the database and fails the check outright.

\paragraph{Repetitive dead loops.} After an error the agent re-issues the identical call
rather than changing strategy, until the step budget is exhausted.

Two of these form a single cascade rather than separate problems: a mis-hearing produces a
wrong argument, the wrong argument a tool error, the tool error a retry of the same call, and the loop burns the
remaining steps until the episode times out at reward zero --- which an outcome-only view sees
as one undifferentiated failure. Audio agentic capability does not follow from audio
understanding.

Every link in the cascade has a reward channel, which makes the diagnosis a training plan.
Mis-heard values surface as failed tool calls, which the per-turn reward penalises while
crediting calls that succeed, so two rollouts that both fail are still ranked by how many
calls landed. Over-action is penalised by the outcome check itself, since the unauthorised
write fails the database comparison. Dead loops are attacked directly, since each repeated
failing call draws its own negative signal instead of being amortised into one terminal zero.

\subsection{Training in SpeechGym}\label{sec:training}

\begin{table}[t]
\centering
\caption{\textbf{The process reward keeps groups informative.} At the base success rates of audio agentic tasks almost every group is all-zero under outcome-only reward; per-turn shaping keeps
nearly all of them informative. Statistics are over the training rollouts of the two reward
configurations on the same task suite.}
\label{tab:groups}
\small
\begin{tabular}{@{}lcc@{}}
\toprule
 & \textbf{Process (ours)} & \textbf{Outcome-only} \\
\midrule
Groups carrying gradient ($\sigma_i > 0$) & \textbf{99.6\%} & 16\% \\
Groups skipped ($\sigma_i = 0$)           & 0.4\%           & 84\% \\
\bottomrule
\end{tabular}
\end{table}

\paragraph{The obstacle is gradient starvation, not optimisation.} At these base success rates
almost every outcome-only group is $K$ identical failures and carries no gradient
(\cref{sec:process}): only 16\% of groups carry gradient, and the other 84\% are discarded
after being generated, their rollout compute spent on trajectories that never touch the
weights (\cref{tab:groups}). Under per-turn shaping, 99.6\% of groups carry gradient,
since two rollouts that both fail still differ in how many tool calls they got right.

Outcome-only GRPO still trains, but four fifths of its rollout budget produces no gradient at
all. Per-turn shaping recovers that budget in the classical manner of reward shaping
\citep{ng1999shaping}; note that our bonus is not potential-based, so it is the terminal
criterion --- \ttb{}'s unmodified check (\cref{sec:reward}) --- and not the objective that is
preserved unchanged.

All of this shares SpeechGym's own audio, user simulator and rollout machinery with training,
so a policy could improve on it by fitting the environment.

\subsection{Cross-pipeline transfer to \tv{}}\label{sec:transfer}

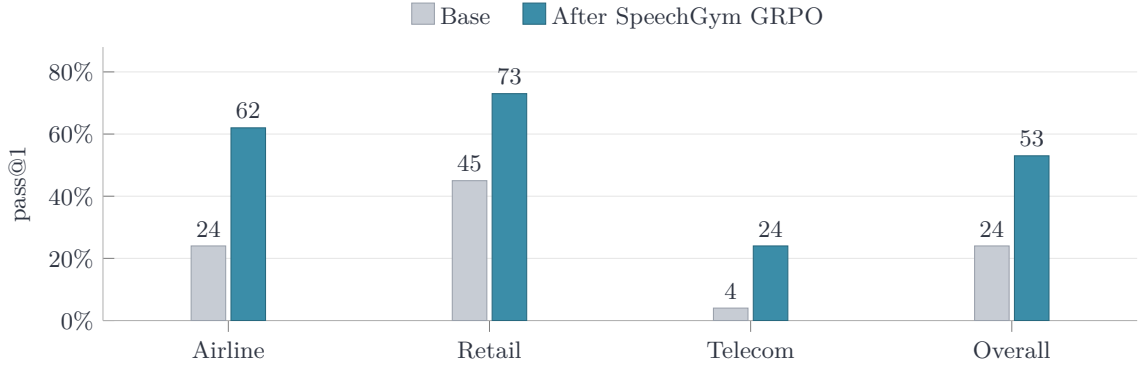
\begin{figure}[t]
\centering
\definecolor{tvGray}{HTML}{C7CCD4}\definecolor{tvGrayE}{HTML}{9AA1AC}
\definecolor{tvAmb}{HTML}{3B8DA8}\definecolor{tvAmbE}{HTML}{27697F}
\definecolor{tvTxt}{HTML}{2F3542}
\begin{tikzpicture}
  \begin{axis}[
      ybar,
      bar width=13pt,
      width=0.92\linewidth, height=5.2cm,
      symbolic x coords={Airline, Retail, Telecom, Overall},
      xtick=data,
      ymin=0, ymax=88,
      ytick={0,20,40,60,80},
      yticklabel={\pgfmathprintnumber{\tick}\%},
      ylabel={pass@1},
      ylabel style={font=\small, text=tvTxt},
      tick label style={font=\small, text=tvTxt},
      axis lines*=left,
      axis line style={black!30},
      ymajorgrids=true,
      grid style={black!10, thin},
      enlarge x limits=0.16,
      nodes near coords,
      every node near coord/.append style={font=\small, text=tvTxt},
      legend style={at={(0.5,1.03)}, anchor=south, draw=none, fill=none,
                    legend columns=2, font=\small, text=tvTxt,
                    /tikz/every even column/.append style={column sep=10pt}},
      legend image code/.code={\draw[#1] (0cm,-0.09cm) rectangle (0.30cm,0.19cm);},
    ]
    \addplot[fill=tvGray, draw=tvGrayE] coordinates {(Airline,24) (Retail,45) (Telecom,4) (Overall,24)};
    \addplot[fill=tvAmb, draw=tvAmbE] coordinates {(Airline,62) (Retail,73) (Telecom,24) (Overall,53)};
    \legend{Base, After SpeechGym GRPO}
  \end{axis}
\end{tikzpicture}
\caption{\textbf{Cross-pipeline transfer to \tv{}} \citep{ray2026tauvoice}, scored by
\tv{}'s own database-and-communication check. The SpeechGym-trained agent is run inside
\tv{}'s independently implemented harness --- its own cascaded user simulator, acoustics
and scoring code --- with \emph{no further tuning}. Overall pass@1 more than doubles, from $24\%$ to $53\%$,
with gains in all three domains --- Airline $24\%$ to $62\%$, Retail $45\%$ to $73\%$,
Telecom $4\%$ to $24\%$ --- and the largest relative gain in Telecom ($6\times$). Both bars in a group come from the same \tv{}
pipeline --- same tasks, same user simulator, same acoustics, same grader --- at a single
attempt; only the agent's weights differ. 95\% confidence intervals do not overlap in any
domain.}
\label{fig:tauvoice}
\end{figure}

\begin{figure}[t]
\centering
\definecolor{lbGray}{HTML}{C7CCD4}\definecolor{lbGrayE}{HTML}{9AA1AC}
\definecolor{lbAmb}{HTML}{3B8DA8}\definecolor{lbAmbE}{HTML}{27697F}
\definecolor{lbPale}{HTML}{D8E8EE}\definecolor{lbTxt}{HTML}{2F3542}

\begin{tikzpicture}
  \begin{axis}[
      xbar, bar width=9pt, bar shift=0pt,
      width=0.80\linewidth, height=6.0cm,
      xmin=0, xmax=80,
      xtick={0,20,40,60}, xticklabel={\pgfmathprintnumber{\tick}\%},
      xlabel={average pass@1}, xlabel style={font=\small, text=lbTxt},
      symbolic y coords={base,gemini,gpt1,casc,gpt2,ours,grok},
      ytick={base,gemini,gpt1,casc,gpt2,ours,grok},
      yticklabels={{\textbf{Base} (no RL)},Gemini-Live-2.5,GPT-Realtime-1.0,
                   Cascaded baseline,GPT-Realtime-2,{\textbf{Ours (SpeechGym)}},Grok-Voice (xAI)},
      y tick label style={font=\small, text=lbTxt},
      tick label style={font=\small, text=lbTxt},
      axis lines*=left, axis line style={black!30},
      xmajorgrids, grid style={black!10},
      tick style={draw=none},
      enlarge y limits=0.10,
      nodes near coords, point meta=x,
      every node near coord/.append style={font=\small, text=lbTxt, anchor=west, xshift=1pt,
                                           /pgf/number format/precision=0, /pgf/number format/fixed},
    ]
    \addplot[fill=lbGray, draw=lbGrayE] coordinates {(26,gemini) (30,gpt1) (31,casc) (51,gpt2) (67,grok)};
    \addplot[fill=lbPale, draw=lbAmbE]  coordinates {(24,base)};
    \addplot[fill=lbAmb,  draw=lbAmbE]  coordinates {(53,ours)};
  \end{axis}
\end{tikzpicture}

\caption{\textbf{\tv{} standing.} Our trained open-weights 30B model against commercial voice
agents on the same benchmark. Training moves the \emph{same} model from last place to second,
ahead of the cascaded baseline, both GPT-Realtime versions and Gemini-Live, with only
Grok-Voice ranking higher. Scores for the other systems are as reported on the benchmark by
their providers; we did not train or re-run them.}
\label{fig:leaderboard}
\end{figure}
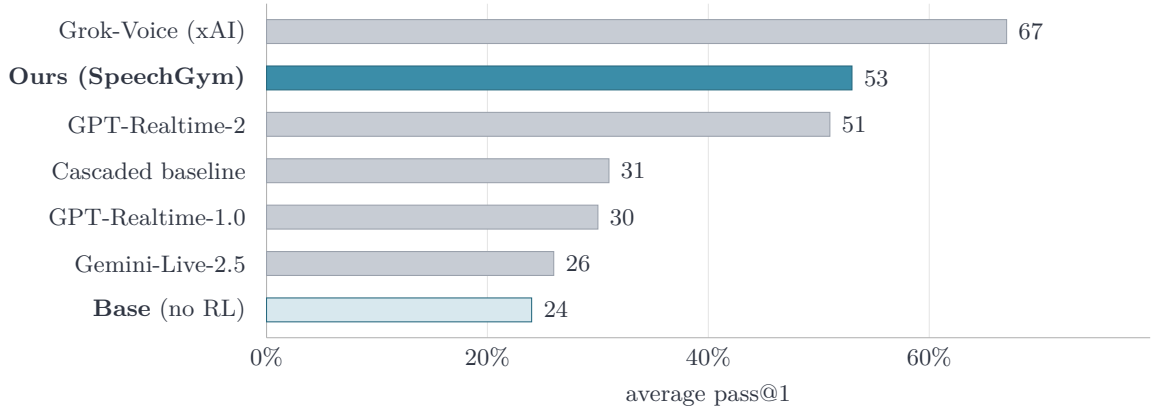

Transferable skill, or overfitted environment? We take the trained checkpoint, apply \emph{no
further tuning}, and run it inside \tv{} \citep{ray2026tauvoice}, an evaluation-only
voice-agent benchmark on the same \ttb{} task set, implemented independently.

\paragraph{What is held fixed, and what is not.} Holding task content fixed isolates the
variable we care about, the audio and interaction pipeline. This is not answer-level
memorisation: a \ttb{} task is a scenario, not a fixed dialogue. The conversation does not
exist until it is generated, turn by turn, by a user simulator that reveals information only
when asked and reacts to whatever the agent says; there is no transcript to replay. Succeeding
means executing the workflow --- eliciting the right identifiers, calling the right tools with
the right arguments, confirming the outcome --- against a user that behaves differently every
time.

The pipeline itself is independent throughout. \tv{} drives its user side with a cascade of
commercial APIs (ASR, a user LLM and TTS), not our model-native simulator; its audio comes
from different TTS voices over a narrowband, telephony-grade channel, against the clean
self-play audio the agent trained on; and it scores with its own harness. Acoustics, user
policy and grader all change at once.

\paragraph{Results.} \Cref{fig:tauvoice} shows overall pass@1 rising from 24\% to 53\%, more
than doubling, with gains in every domain and the largest relative gain in Telecom
($6\times$, from a base of 4\%). The 95\% confidence intervals do not overlap in any domain.
The only thing that differs is the agent's weights.

\Cref{fig:leaderboard} gives context: the same open 30B model moves from last place to second
among the systems reported on this benchmark. We keep the claim calibrated --- these are
systems we did not build, train or re-run, evaluated by their providers' own deployed stacks.
What we can say is that one open-weights model, trained locally at no API cost, reaches this
position under the same check, and that the change came from RL rather than scale, since the
base of the same model sits at the bottom.

\subsection{Why it improves: mechanistic checks}\label{sec:mechanism}

\begin{table}[t]
\centering
\caption{\textbf{Behavioural change on \tv{}}, computed from the raw trajectories of both
models on the same task set. Every behaviour we tracked moves in the intended
direction: wrong writes, i.e.\ over-action, drop by more than half, dead-loops fall by roughly
two thirds, and the agent recovers from a mis-hearing far more often.}
\label{tab:behaviour}
\small
\begin{tabular}{@{}lccc@{}}
\toprule
\textbf{Behaviour} (\% of tasks) & \textbf{Base} & \textbf{After GRPO} & \textbf{$\Delta$} \\
\midrule
Wrong writes (over-action), lower better        & 23\% & 10\% & $-13$ \\
Dead-loops when stuck, lower better             & 14\% & 5\%  & $-9$ \\
Recovers after a mis-hearing, higher better     & 42\% & 62\% & $+20$ \\
\bottomrule
\end{tabular}
\end{table}

\begin{table}[t]
\centering
\caption{\textbf{Higher success at lower cost} (\tv{}). Success more than doubles while
turns and tokens both fall, which rules out the two standard ways an RL agent can inflate
a success metric: taking more turns until something works, or stalling. For reference the
cascaded baseline needs 31.4 turns to reach 31\% pass@1.}
\label{tab:efficiency}
\small
\begin{tabular}{@{}lccc@{}}
\toprule
\textbf{Metric} & \textbf{Base} & \textbf{After GRPO} & \textbf{$\Delta$} \\
\midrule
pass@1                       & 24\%   & 53\%   & more than doubles \\
Average agent turns          & 26     & 24     & $-8\%$ \\
Average tokens per task      & 51{,}195 & 48{,}398 & $-5\%$ \\
Turns on the tasks it fixes  & 23.4   & 17.5   & $-25\%$ \\
\bottomrule
\end{tabular}
\end{table}

A jump from 24\% to 53\% invites the suspicion that a metric was gamed rather than a task
solved. Two
checks, either of which could have falsified the result.

\paragraph{(a) Did the diagnosed failures go away?} Had the gain come from elsewhere than the
cascade of \cref{sec:failures}, the diagnosed rates would be roughly unchanged. We annotated
the raw \tv{} trajectories of both models on the same task set for the behaviours that the cascade
identifies (\cref{tab:behaviour}). Unauthorised writes drop from 23\% to 10\% of tasks, dead
loops when stuck from 14\% to 5\%, and recovery after a mis-hearing rises from 42\% to 62\%.
The failures the environment was built to expose are the failures that move.

\paragraph{(b) Did it buy success with more interaction?} Inflating a success rate by spending
more of the episode budget --- retrying, or stalling until the user concedes --- predicts
turns and tokens rising with pass@1. They fall (\cref{tab:efficiency}).
Agent turns go from 26 to 24 and tokens per task from 51,195 to 48,398 while pass@1 more than
doubles; on the tasks the trained model fixes it is 25\% more concise than the base was
(23.4 to 17.5 turns). Against the cascaded baseline: 24 turns at 53\% pass@1 versus 31.4 turns
at 31\%. Success rising while compute falls is the opposite of length-based reward hacking.

\paragraph{Strategies we did not design.} The annotations also record repair strategies no
part of the system specifies: asking the caller to spell a name out, retrying a lookup with a
corrected spelling, switching lookup key when the first fails. Nothing in the reward mentions
spelling, retries or lookup keys --- it scores task completion and tool-call success --- so RL
found them, consistent with their raising the chance of completing a task over an unreliable
channel.

\section{Conclusion}\label{sec:conclusion}

SpeechGym is, to our knowledge, the first audio-native agentic environment that both
evaluates and trains voice agents end to end in speech, with a local omni-modal user in the
loop in place of an API cascade. Using it, we find that the dominant failures of a voice
agent are perceptual --- slot values misheard from a waveform, and the cascade of failed
calls and repetition loops that follows --- rather than reasoning deficits, so that audio
comprehension and audio \emph{agency} are distinct capabilities; that GRPO with a per-turn
process reward closes much of the resulting gap; and that the skills so acquired transfer
without further tuning to an independently implemented benchmark, more than doubling pass@1
there from 24\% to 53\%. The broader point is one of framing: the text-to-audio gap has
until now been an open measurement, and an environment that closes the loop recasts it as an
optimisation problem with an objective, a gradient and a stopping criterion --- and because
the interface is a standard reset/step/reward loop over text agentic domains, any open
omni-modal model drops into it, in the way standardised environments have served other
capabilities \citep{bellemare2013ale,gdi,lbc}.

\section*{Broader Impact}
Every task in SpeechGym runs against a synthetic relational database populated with
fictional users, so no real personal information is processed in training or evaluation, and
the reward is defined entirely by task completion --- the correct final database state and
the correct information communicated --- with no term rewarding persuasion, pressure or any
other manipulation of the simulated caller. More capable voice agents nevertheless carry
deployment risks a simulated gym does not address, and responsible deployment requires
safeguards outside its scope: content filtering, explicit user consent for recorded or
synthesised speech, and reliable escalation to a human. Because the environment tracks
policy-relevant behaviour directly, failures such as unauthorised writes become
\emph{measurable} quantities that respond to training (\cref{tab:behaviour}), and
measurability is a prerequisite for control.

\bibliographystyle{plainnat}
\bibliography{main}

\end{document}